\documentclass[
twocolumn,
amsmath,
amssymb,
aps,
]{openjournal}

\usepackage{graphicx}
\usepackage{dcolumn}
\usepackage{bm}
\usepackage{soul} 
\usepackage[colorlinks=True]{hyperref}
\usepackage{natbib}
\usepackage[bottom]{footmisc}
\usepackage{booktabs}
\usepackage{amsmath}
\usepackage[mathlines]{lineno}

\begin{document}

\title{Comparing Mass Mapping Reconstruction Methods with Minkowski Functionals}

\author{Nisha Grewal$^1 \ast$, Joe Zuntz$^1$, Tilman Tr\"oster$^2$}
\affiliation{
    $^1$Institute for Astronomy, University of Edinburgh, Royal Observatory, Blackford Hill, Edinburgh, EH9 3HJ, UK\\
    $^2$Eidgenoessische Technische Hochschule Z\"urich, R\"amistrasse 101, 8092 Z\"urich, Switzerland 
}

\email{$\ast$ nisha.grewal@ed.ac.uk}

\date{\today}

\begin{abstract}
        Using higher-order statistics to capture cosmological information from weak lensing surveys often requires a transformation of observed shear to a measurement of the convergence signal.  This inverse problem is complicated by noise and boundary effects, and various reconstruction methods have been developed to implement the process. Here we evaluate the retention of signal information of four such methods: Kaiser-Squires, Wiener filter, \texttt{DarkMappy}, and \texttt{DeepMass}. We use the higher order statistics \textit{Minkowski functionals} to determine which method best reconstructs the original convergence with efficiency and precision. We find \texttt{DeepMass} produces the tightest constraints on cosmological parameters, while Kaiser-Squires, Wiener filter, and \texttt{DarkMappy} are similar at a smoothing scale of 3.5 arcmin. We also study the MF inaccuracy caused by inappropriate training sets in the \texttt{DeepMass} method and find it to be large compared to the errors, underlining the importance of selecting appropriate training cosmologies.
\end{abstract}

\maketitle


\section{\label{sec:level1}Introduction}

Weak gravitational lensing uses the distortion of the shapes and/or size of galaxies as a probe of the matter distribution along the line of sight. Maps quantifying lensing can be generated by measuring the observed shapes of galaxies, utilizing this shear effect from gravitational lensing \citep{hsc_shear2,hsc_shear,kids_shear2,kids_shear,des_shear}. Statistics measured from weak gravitational lensing are effective probes of the evolution of large-scale structure. Galaxy surveys, such as the Dark Energy Survey\footnote{\url{https://www.darkenergysurvey.org/}} \citep[DES; ][]{des_survey}, Hyper Supreme-Cam\footnote{\url{https://www.naoj.org/Projects/HSC/}} \citep[HSC; ][]{hsc_survey} survey, and Kilo-Degree Survey\footnote{\url{http://kids.strw.leidenuniv.nl/}} \citep[KiDS; ][]{kids_survey}, constructed detailed galaxy catalogues of tens to hundreds of millions of observed galaxies \citep{des_cat,kids_cat,hsc_cat}. The Legacy Survey Space and Time at the Rubin Observatory\footnote{\url{https://www.lsst.org/}} \citep[LSST; ][]{lsst_book,lsst_desc} and the space telescopes Euclid\footnote{\url{https://www.euclid-ec.org/}} \citep{euclid} and Nancy Grace Roman\footnote{\url{https://roman.gsfc.nasa.gov/}} \citep{roman} will go further in depth and area, and so improve our understanding of dark matter and dark energy.

Weak lensing causes source galaxies shapes to be magnified by the convergence field and distorted by the shear field. While only the shear signal can be directly inferred from galaxy shape observations, convergence can conveniently be represented as an integral of the projected matter density along the line of sight and can be derived from shear. Without noise or boundary effects, the observed shear field can be used to measure the convergence field trivially, but this is not the case for real data. We need to solve an inverse problem, working backwards from observational shear measurements to infer the underlying convergence in the presence of these effects in real data, which becomes complex.

Various non-parametric mass mapping reconstruction methods have been developed to address this problem and take into account noise and masks in the data to solve the inverse problem. Linear approaches like Kaiser Squires (KS) \citep{kaisersquires93}, Wiener filter (WF) \citep{wiener_1994}, \citet{linear_recon_1}, and \citet{linear_recon_2} and non-linear approaches like \texttt{GLIMPSE} \citep{glimpse}, Glimpse2D \citep{glimpse2d}, KS+ \citep{ks_plus}, \texttt{DeepMass} \citep{deepmass}, MCAlens \citep{mcalens}, \texttt{DarkMappy} \citep{darkmappy}, \texttt{KaRMMa} \citep{karmma}, \texttt{DeepPosterior} \citep{remy_probabilistic}, SKS+ \citep{sks_plus}, and others \citep{nonlinear_recon_1,wiener2012,nonlinear_recon_2,nonlinear_recon_3,nonlinear_recon_4,wiener2018,bayesian2} have all been shown to be effective. The original method, Kaiser-Squires (KS) is a linear inversion of the shear map into a convergence map, but it cannot take noise or boundary effects into account \citep{kaisersquires93}. Wiener filtering is also a linear filter, but unlike KS, it can transform noisy data well \citep{wiener_1994}. Sparsity methods like \texttt{GLIPMSE} and \texttt{DarkMappy} are non-linear and able to reconstruct information in the non-Gaussian regime \citep{glimpse,darkmappy}. Some methods like \texttt{DarkMappy}  and \texttt{DeepPosterior} are able to approximate the uncertainty that comes from reconstruction. Other methods including \texttt{DeepPosterior} have employed machine learning as a useful tool in denoising convergence signal \citep{deepmass,hsc_gan,remy_probabilistic}. In this paper we explore four reconstruction methods: Kaiser Squires, Wiener filter, \texttt{DarkMappy}, and \texttt{DeepMass}.

Measuring statistics from the best reconstructed convergence maps is expected to lead to tighter constraints on cosmological parameters. Two-point statistics, which describe the correlation between pairs of data points, can constrain these parameters with high precision \citep{kids_cosmo,hsc_cosmo,des_cosmo}. However, statistics beyond two-point are required to fully probe non-Gaussian structure on small scales. Higher order statistics including N-point such as the one-point probability distribution function \citep{liu2019constraining,barthelemy2020nulling,thiele2020accurate,boyle2021nuw} and three-point correlation functions \citep{3ptcorr1,3ptcorr2,3ptcorr3}, moments \citep{moment1,moment2,moment3,moment4,moment6,moment5,moment7,gatti_moments_2021,moment8}, topological descriptors such as Betti numbers \citep{feldbrugge2019stochastic,betti1,parroni2021higher} and peak counts \citep{marian2009cosmology,kratochvil_2010,liu_2015,Kacprzak_2016,KIDS_peak,martinet_2018,moment6,ajani_2020,harnois2021cosmic,zurcher2021}, scattering transform coefficients \citep{cheng2020new,cheng2021weak,gatti2023dark}, and field level inference \citep{bayesian2,boruah2022map} can measure information in the non-Gaussian regime in weak lensing maps. 

Another higher-order statistic, Minkowski functionals (MFs) are morphological descriptors that describe the topology of a continuous field \citep{minkowski}. MFs are unbiased, have low variance, and can have additional resilience to some systematic uncertainties, making them effective probes of the underlying dark matter distribution \citep{non-Gaus}. MFs have been applied to Cosmic Microwave Background data \citep{cmb_mf,cmb_mink1,cmb_mink2,cmb_mink3,cmb_mink4,cmb_mink5,cmb_mink6}, 2D convergence maps \citep{first_mf,kratochvil,Petri,vicinanza,deep_minkowski,grewal,euclid_metric}, 2D density fields \citep{grewal}, and 3D density fields from spectroscopic data \citep{3Dclust_1,3Dclust_2,3Dclust_3,3Dclust_4,mink_neutrinos}. Like other higher-order statistics, MFs produce tighter constraints on cosmological parameters because they are sensitive to small-scale structure.  

In this paper, we use MFs as a test of different convergence reconstruction methods, using them to quantify the information (particularly on small scales) that the methods can reconstruct in maps. We compare the four different mass mapping methods KS, WF, \texttt{DarkMappy}, and \texttt{DeepMass}, applying them to simulated images and using MFs to constrain $w$CDM parameters and thus computing figures of merit to describe method power. 

In Section \ref{section:Formalism} we review weak lensing formalism and the mass mapping reconstruction methods, in Section \ref{section:Methodology} we describe the simulations, observables, and evaluation metrics, in Section \ref{section:Results} we compare the cosmological constraints from the four reconstruction methods, and we conclude in Section \ref{section:Conclusion}.

\section{Formalism}
\label{section:Formalism}

\subsection{Convergence and Shear}

The magnification and distortion of a lensed galaxy image is represented to first order by the Jacobian matrix

\begin{equation}
    \mathcal{A}(\boldsymbol{\theta}) = \left( \delta_{ij} - \frac{\partial^2 \psi(\boldsymbol{\theta})}{\partial \theta_i \partial \theta_j} \right) =
    \begin{pmatrix}
    1 - \kappa - \gamma_1 & -\gamma_2 \\
    -\gamma_2 & 1 - \kappa + \gamma_1
    \end{pmatrix},
    \label{jacobian}
\end{equation}
where $\theta$ is the observed angular position, $\psi$ is the lensing potential, the shear $\gamma \equiv \gamma_1 + i\gamma_2$, and convergence is $\kappa$ \citep{weaklens}. Convergence is a measure of the matter distribution along the line of sight. It is given by a projection of the overdensity $\delta$:

\begin{equation}
    \kappa(\mathbf{\hat{\theta}}) = \frac{3\mathrm{H}_0^2\Omega_\mathrm{m}}{2c^2} \int_0^{\chi_\mathrm{lim}} d\chi \frac{q(\chi)}{a(\chi)}f_K(\chi)\delta(f_K(\chi)\mathbf{\hat{\theta}},\chi),
\end{equation}
where $\mathrm{H}_0$ is the Hubble constant, $\Omega_\mathrm{m}$ is the matter density, $c$ is the speed of light, $\chi$ is comoving distance, $\chi_\mathrm{lim}$ is the horizon distance, $a$ is the scale factor, and the lens efficiency $q$ is defined as

\begin{equation}
    q(\chi) = \int_0^{\chi_\mathrm{lim}}n(\chi')d\chi'\frac{f_K(\chi'-\chi)}{f_K(\chi')}.
\end{equation}
$n(\chi)\mathrm{d}\chi$ is the source galaxy distribution and $f_K$ is the comoving angular distance \citep{Kilbinger_2015}. 

While convergence is harder to observe directly, shear can be measured from galaxy shape observations. The convergence $\kappa$ is related to the shear $\gamma$ via the lensing potential $\psi$. Rearranging the following equations for $\kappa$ and $\gamma$ and solving for $\psi$,

\begin{equation}
    \kappa = \frac{1}{2}\Delta\psi,
\end{equation}

\begin{equation}
    \gamma_1 = \frac{1}{2}\left(\partial_1^2\psi-\partial_2^2\psi\right);     
    \gamma_2 = \partial_1\partial_2\psi,
\end{equation}
and then performing a Fourier transform in the flat sky regime, the convergence can be expressed as

\begin{equation}
    \Tilde{\kappa} = \frac{k_1^2 - k_2^2}{k^2}\Tilde{\gamma_1} + \frac{2k_1k_2}{k^2}\Tilde{\gamma_2}.
    \label{eq:convergence}
\end{equation}
$k_1$ and $k_2$ are spatial frequencies in Fourier space ($k^2 = k_1^2 + k_2^2$), and $\Tilde{\gamma_1}$ and $\Tilde{\gamma_2}$ are the two orthogonal components of shear distortion \citep{remy_probabilistic}. As the expectation value of matter overdensity $\delta$ is zero, the first moment of shear and convergence will also be zero. Known as mass sheet degeneracy, this feature means that at $k$ = 0, the relationship between convergence and shear is undefined. It is important to note that this transformation is only optimal when performed on pure signal in the absence of noise and missing data.

In the presence of noise and gaps from masking, we must go beyond this simple treatment. In this more realistic regime, various algorithms attempt the inversion from shear to convergence in different ways. We review several of these methods here.

\subsection{Kaiser-Squires}

Kaiser-Squires uses a direct linear inversion of the shear in Fourier space to obtain the reconstructed convergence field \citep{kaisersquires93}. Scalar perturbations in matter density are captured by the E mode gradient-like patterns, and the tensor perturbation counterparts are captured by the curl-like B mode. The shear-convergence relation can be expanded to give both the E and B modes:

\begin{equation}
    \Tilde{\kappa} = \Tilde{\kappa_E}+i\Tilde{\kappa_B} = \left( \frac{k_1^2 - k_2^2}{k^2} + i\frac{2k_1 k_2}{k^2} \right) (\Tilde{\gamma_1}+i\Tilde{\gamma_2}).
\end{equation}

We can define the forward model in Fourier space $\mathbf{P}$ as

\begin{equation}
    \textbf{\textrm{P}} = \left( \frac{k_1^2 - k_2^2}{k^2} + i\frac{2k_1 k_2}{k^2} \right)
    \label{eq:P_kappa}
\end{equation}
\citep{kaisersquires93}. We note that we do not expect weak lensing to generate B modes, which come from systematics and higher-order lensing effects.

KS disregards noise and boundary effects, which means it requires continuous fields to reconstruct maps with masks \citep{moment1}. In common with other approaches, we use a Gaussian kernel to fill in the gaps and smooth the field, though this leads to loss of structure on small scales and suppresses peaks in the convergence \citep{Jeffrey_2018}. This also leads to leakage of the E and B modes of the reconstructed convergence field \citep{remy_probabilistic}. 

\subsection{Wiener filter}

A Wiener filter (WF) is a linear inversion of signal and noise together, aiming to suppress the latter. The WF method uses the expected power spectrum of the convergence field at a fiducial cosmology and of the noise \citep{wiener2012}. In Fourier space:

\begin{equation}
    \Tilde{\kappa} = \textbf{SP}^\dag \left[ \textbf{\textrm{PSP}}^\dag + \textbf{\textrm{N}} \right]^{-1} \Tilde{\gamma},
    \label{eq:wf}
\end{equation}
where \textbf{S} is the covariance matrix of power spectrum signal and \textbf{N} is the noise covariance matrix \citep{wiener_1994,wiener_1995}. However, in filtering out the noise, WF smooths out small-scales features in the map more generally. While WF can reconstruct the field with precision on large scales, it does not typically retain small-scale information in convergence maps as well \citep{wiener2012,wiener2018}.

WF balances noise reduction and signal preservation. Underestimating the noise level causes WF to misinterpret noise as signal, resulting in a loss of accuracy; overestimating the noise causes WF to suppress signal as noise, affecting the reconstruction precision as the noise is not reduced properly \citep{wiener2012,wiener2018}.

\subsection{\textnormal{\texttt{DarkMappy}}}

\texttt{DarkMappy} was primarily designed for cluster analysis; here we apply the method to a wide-field convergence. \texttt{DarkMappy} models the convergence field using wavelets \citep{darkmappy}. It uses a sparsity approach to minimise the number of wavelets used and varies their coefficients in a hierarchical Bayesian approach, which enables an approximation of the signal uncertainty that comes from reconstruction. As it is a non-linear approach, we expect it to measure small-scale effects smoothed out by KS and WF.

A wavelet dictionary is a collection of wavelets, which are oscillatory functions with zero mean that are localised in Fourier and real space \citep{wavelet}. A wavelet dictionary $\mathbf{\Psi}$ can sparsely represent the signal $\kappa$:

\begin{equation}
    \kappa = \mathbf{\Psi}\alpha,
\end{equation}
where only a subset of non-zero coefficients $\alpha$ is needed \citep{glimpse}. 

Unlike the previous methods, \texttt{DarkMappy} fits the data in real space. The forward model operation in Eq.~\eqref{eq:P_kappa} is transformed into a measurement operator $\mathbf{\Phi}$ using a Fourier transform $\mathbf{F}$:

\begin{equation}
    \mathbf{\Phi} = \mathbf{F}^{-1}\mathbf{PF}.
\end{equation}
$\mathbf{\Phi}$ maps convergence onto shear in the measurement equation, where the total shear is then given by:

\begin{equation}
    \gamma = \mathbf{\Phi}\kappa + n,
\end{equation}
where $n$ is the noise contaminating the signal. Using a wavelet-based, sparsity-promoting, $l_1$-norm prior, the \texttt{DarkMappy} model for the reconstructed map is then given by the \textit{maximum a posteriori} solution: 

\begin{equation}
    \kappa^{\mathrm{map}} = \underset{\kappa}{\mathrm{argmin}} \biggl\{ \mu\parallel\mathbf{\Psi}^\dag\kappa\parallel_1 
    + \frac{1}{2}\parallel \Sigma^{-\frac{1}{2}} (\mathbf{\Phi}\kappa - \gamma)\parallel_2^2 \biggr\}.
\end{equation}
The first term is the prior, which imposes sparsity, and the second term is the likelihood of the data. $\mu$ regularises the weighting between the two terms and $\Sigma$ is the covariance of the measured shear \citep{darkmappy}.

The search process for the MAP uses \textit{knock-out} hypothesis testing of the posterior to differentiate between original signal and reconstruction effects \citep{knockout}. This algorithm tests whether features (e.g. peaks) are significant enough to be included in the MAP model using a statistical test that compares $\kappa$ models that include them to those using $\kappa$ smoothed by surrounding pixels.

\subsection{\textnormal{\texttt{DeepMass}}}

\texttt{DeepMass} is a convolutional neural network (CNN) that aims to denoise noisy maps. The network learns parameters $\Theta$ in convolution layers to reconstruct convergence maps from noisy shear maps:

\begin{equation}
    \kappa = \mathcal{F}_\Theta(\gamma),
\end{equation}
where $\mathcal{F}_\Theta(\gamma)$ is the posterior estimate of convergence \citep{deepmass}. Convergence and shear training data is used to minimise a mean-square-error (MSE) cost function $J$ as the network finds $\Theta$:

\begin{equation}
    J(\Theta) = \parallel\mathcal{F}_\Theta(\gamma) - \kappa\parallel_2^2.
\end{equation}

Conceptually, the `true' output noise-free convergence maps $\boldsymbol{\kappa}$ are drawn from a prior distribution $\mathrm{P}(\kappa)$, and the corresponding input noisy shear maps are drawn from the likelihood $\mathrm{P}(\gamma|\kappa)$ \citep{deepmass}. The reconstructed convergence $\kappa$ can then be approximated as

\begin{equation}
    \kappa = \mathcal{F}_\Theta(\boldsymbol{\gamma}) = \int \kappa \mathrm{P}(\kappa|\gamma) \textrm{d}\kappa.
\end{equation}

In practice, the network is trained on pairs of truth convergence maps and noisy WF convergence maps, which are reconstructed from shear maps \citep{deepmass}. The model is two-dimensional, so that multiple models need to be built for data with more than one redshift bin. \texttt{DeepMass} utilises a U-Net architecture with a contracting path (encoder), reducing the dimensions of the input data while extracting high-level features, and an expanding path (decoder), which complements the contracting path by restoring the full spatial dimension and generating a high-resolution output \citep{unet}. It also uses average pooling, where each region of the input data is replaced with its mean value. As the resolution decreases, the model captures more comprehensive physical features in the convergence map by considering a wider area and reducing the dimensionality of the data \citep{avg_pooling}. While \texttt{DeepMass} can measure small-scale structure in convergence maps, the model does not retrieve uncertainties in the reconstruction \citep{deepmass,remy_probabilistic}. As with all CNNs, \texttt{DeepMass} risks overfitting to map structure and/or creating fake artefacts in the reconstructed signal. 

\subsection{Minkowski functionals}

Minkowski functionals are field integrals that characterise the topological properties of continuous fields \citep{minkowski,non-Gaus}. In this work we measure these mathematical descriptors from a convergence field. MFs measure the properties of \emph{excursion sets} of this field, which are given by the region of the field above a given threshold \citep{deep_minkowski}. The first three functionals quantify area, perimeter, and mean curvature of an excursion set and can be written\footnote{We write curvature integral summary statistics of MFs in Eqs.~\eqref{eq:v0},~\eqref{eq:v1}, and~\eqref{eq:v2}, which are valid for a field with no boundary. See \citet{first_mf} for the most general definitions.} as

\begin{equation}
    V_0(t) = \frac{1}{A} \int\limits_A \Xi(\kappa(\mathbf{x})-t)\mathrm{d}^2 \mathbf{x}, 
    \label{eq:v0}
\end{equation}

\begin{equation}
    V_1(t) = \frac{1}{4A} \int\limits_A \delta(\kappa(\mathbf{x})-t)\sqrt{\kappa_x^2 + \kappa_y^2} \mathrm{d}^2 \mathbf{x} ,
    \label{eq:v1}
\end{equation}

\begin{multline}
    V_2(t) = \frac{1}{2\pi A} \int\limits_A \delta(\kappa(\mathbf{x})-t) \\
    \left( \frac{2\kappa_x \kappa_y \kappa_{xy}-\kappa_x^2 \kappa_{yy}-\kappa_y^2 \kappa_{xx}}{\kappa_x^2 + \kappa_y^2} \right) \mathrm{d}^2 \mathbf{x} ,
    \label{eq:v2}
\end{multline}
where $\mathbf{x}=(x,y)$ is the location in the field, $t$ is a chosen threshold, $A$ is the total area of the map, $x$ and $y$ are flat sky coordinates, $\kappa(\textbf{x})$ is the field value in two dimensions, and $\kappa_x$, $\kappa_y$, $\kappa_{xy}$, $\kappa_{yy}$, and $\kappa_{xx}$ are derivatives of the field \citep{minkowski,Petri}. The derivatives are also evaluated at position $\mathbf{x}$; this is dropped in Eq.~\eqref{eq:v1} and~\eqref{eq:v2} for brevity.

The Heaviside function $\Xi$ in Eq.~\eqref{eq:v0} identifies the area of the field region above the threshold; similarly, the Dirac delta functions in Eq.~\eqref{eq:v1} and~\eqref{eq:v2} select the regions where the height matches the threshold to measure the perimeter and connectivity of the field as the curvature of the boundary, respectively \citep{minkowski}.

\section{Methodology}
\label{section:Methodology}

\subsection{Simulations}

To generate our convergence maps we use C\textsc{osmo}G\textsc{rid}V1, which comprises a suite of full-sky lightcone simulations with nside = 2048 and shells that allow for probing of multiple redshifts, up to z $<$ 3.5 \citep{cosmogrid}. The simulations were run with PKDGRAV3, a high performance self-gravitating astrophysical N-body treecode \citep{potter2016pkdgrav3}. C\textsc{osmo}G\textsc{rid}V1 varies the cosmological parameters $\Omega_\mathrm{m}$, $\sigma_8$, $w_0$, $n_\mathrm{s}$, $\Omega_\mathrm{b}$, and $\mathrm{H}_0$ and baryonic parameters $M^0_\mathrm{c}$ and $\nu$, but uses fixed neutrino masses, each with $m_v$ = 0.02eV. To make realistic lensing maps, baryonic effects are included using a shell-based baryon correction model at the map level. Particle count maps are adjusted by a 2D displacement function measured from halo catalogues \citep{cosmogrid}. Following \citet{bary_mc}, the mass dependence of the gas profile $M_\mathrm{c}$ is broken down into the amplitude $M_\mathrm{c}^0$ and redshift dependence parameter $\nu$: 

\begin{equation}
    M_\mathrm{c} = M_\mathrm{c}^0(1+z)^{\nu}.
    \label{eq:baryon}
\end{equation}

C\textsc{osmo}G\textsc{rid}V1 provides 200 simulations at a fiducial cosmology with different seeds for the initial conditions, accompanied by maps with paired matching seeds with two steps, one above and one below for each of the cosmological parameters, shown in Table \ref{tab:delta_params}. For each cosmology, we cut out 50,000 15 degree $\times$ 15 degree patches with random positions and rotations from the 200 simulations, removing values at the poles to avoid distortion. The resulting convergence maps have resolution 256 $\times$ 256 pixels with pixel size 3.5 arcmin. 

The C\textsc{osmo}G\textsc{rid}V1 convergence maps have only positive values; here we subtract the mean after reconstruction before measuring observables. The mass-sheet degeneracy means the zero-point is arbitrary when going from shear back to convergence.

\begin{table}[h]
  \centering
  \renewcommand{\arraystretch}{1.3}
  \begin{tabular}{cll}
  \hline    
    & Fiducial & $\Delta$ Fiducial \\
    \midrule
    $\Omega_\mathrm{m}$ & 0.26   & $\pm$ 0.01   \\
    $\sigma_8$ & 0.84   & $\pm$ 0.015  \\
    $w_0$      & -1     & $\pm$ 0.05   \\
    $n_\mathrm{s}$      & 0.9649 & $\pm$ 0.02   \\
    $\Omega_\mathrm{b}$ & 0.0493 & $\pm$ 0.001  \\
    $\mathrm{H}_0$      & 67.3   & $\pm$ 2.0    \\
    $M_\mathrm{c}^0$    & 13.82  & $\pm$ 0.1    \\
    $\nu$      & 0.0    & $\pm$ 0.1    \\
    \bottomrule
  \end{tabular}
  \caption{\textup{C\textsc{osmo}G\textsc{rid}V1 Parameter Shift Values. $\mathrm{H}_0$ has units $\mathrm{kms^{-1}Mpc^{-1}}$.}}
  \label{tab:delta_params}
\end{table}

In this work, we do our analysis with 5,000 map patches, a balance between efficiency and precision found by requiring convergence of the KS contours described below, where there is less than a 5\% shift in contours. For \texttt{DeepMass} training and testing, we use 50,000 random patches from 200 maps to avoid overfitting at the given cosmology. However, all observables are only measured from 5,000 maps.

\subsubsection{Redshift}
\begin{figure}[ht]
    \centering
    \includegraphics[width=\columnwidth]{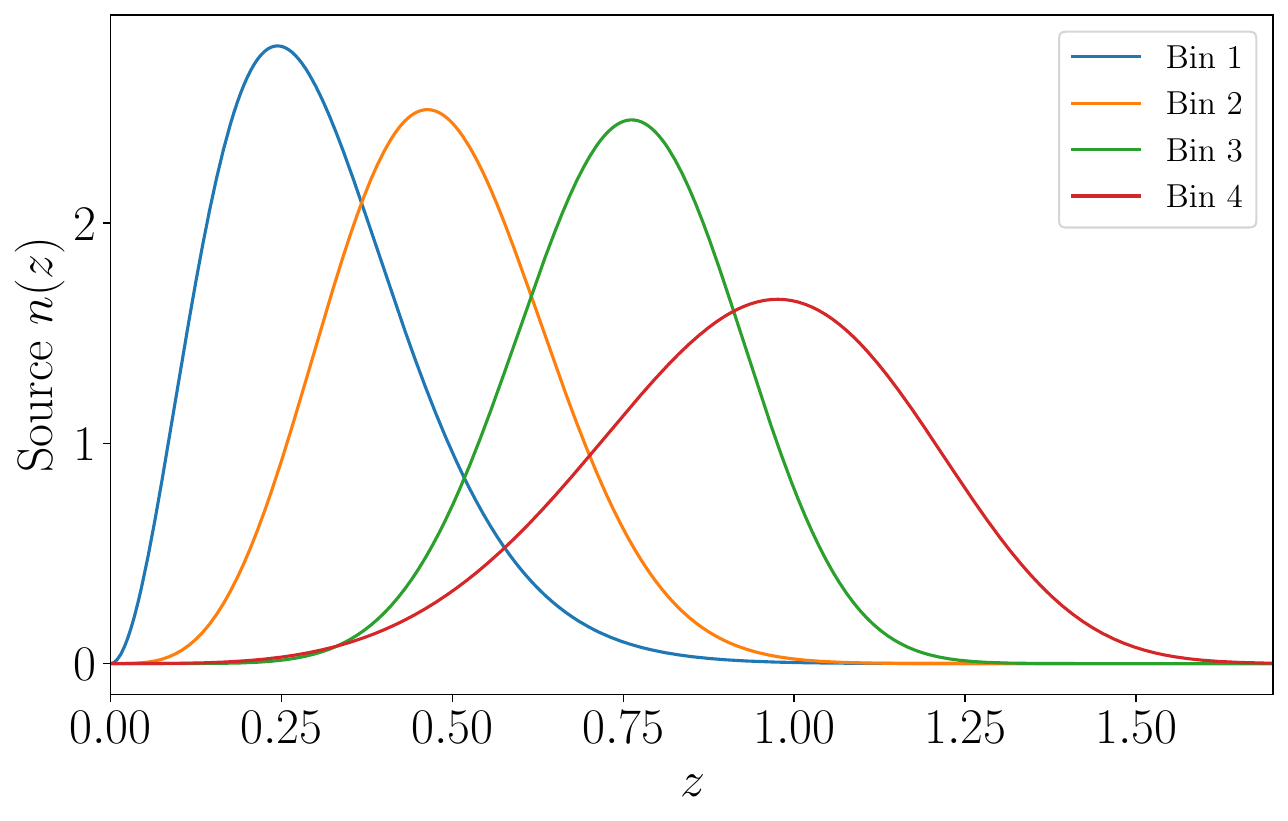}
    \caption{The DES Y3-like $n_z$ redshift distributions from C\textsc{osmo}G\textsc{rid}V1 used in this work \citep{cosmogrid}.}
    \label{fig:redshift}
\end{figure}

In our analysis, we use approximately DES Y3-like tomographic redshifts from the C\textsc{osmo}G\textsc{rid}V1 shells, modelled as a Smail-type distribution: 

\begin{equation}
    n(z) = z^\alpha \mathrm{exp} \left[ - \left( \frac{z}{z_0} \right) ^\beta \right],
\end{equation}
\citep{smail,cosmogrid}. The survey parameter values can be seen in Table \ref{tab:smail} and the final distributions can be seen in Figure \ref{fig:redshift}.

\begin{table}[h]
    \centering
    \renewcommand{\arraystretch}{1.3}
    \begin{tabular}{cccc}
        \hline
        Bin & $\alpha$ & $\beta$ & $z_0$   \\
        \midrule
        1 & 1.99 & 1.44 & 0.20    \\
        2 & 3.46 & 2.34 & 0.39    \\
        3 & 6.03 & 3.60 & 0.66    \\
        4 & 3.53 & 4.49 & 1.03    \\
        \hline
    \end{tabular}
    \caption{\textup{Smail-type distribution parameters for DES Y3-like redshift bins \citep{redshift}}}
    \label{tab:smail}
\end{table}

\subsubsection{Noise and Mask}

At our chosen resolution, our maps are in the low noise regime. We only take shape noise into account, disregarding any other observational effects. To obtain the noise standard deviation per pixel, we use 

\begin{equation}
    \sigma = \frac{\sigma_e}{\sqrt{\frac{n_\mathrm{gal}}{n_\mathrm{bin}}p^2}},
    \label{eq:sigma}
\end{equation}
where the shape dispersion $\sigma_e$ is 0.26 per galaxy, $n_\mathrm{gal}$ is a DES Y3-like number density (10 galaxies/arcmin$^2$), $n_\mathrm{bin}$ is the four DES Y3-like redshift bins, and $p$ is the pixel size in arcmin.  With 256$^2$ total pixels and 15 $\times$ 15 square degree maps, the pixel size $p$ is 3.5 arcmin and the noise level $\sigma$ is 0.047 per pixel. The map noise is computed as an uncorrelated random Gaussian field with mean 0 and $\sigma$ from Eq.~\eqref{eq:sigma}. 

\begin{figure}[ht]
    \centering
    \includegraphics[width=\columnwidth]{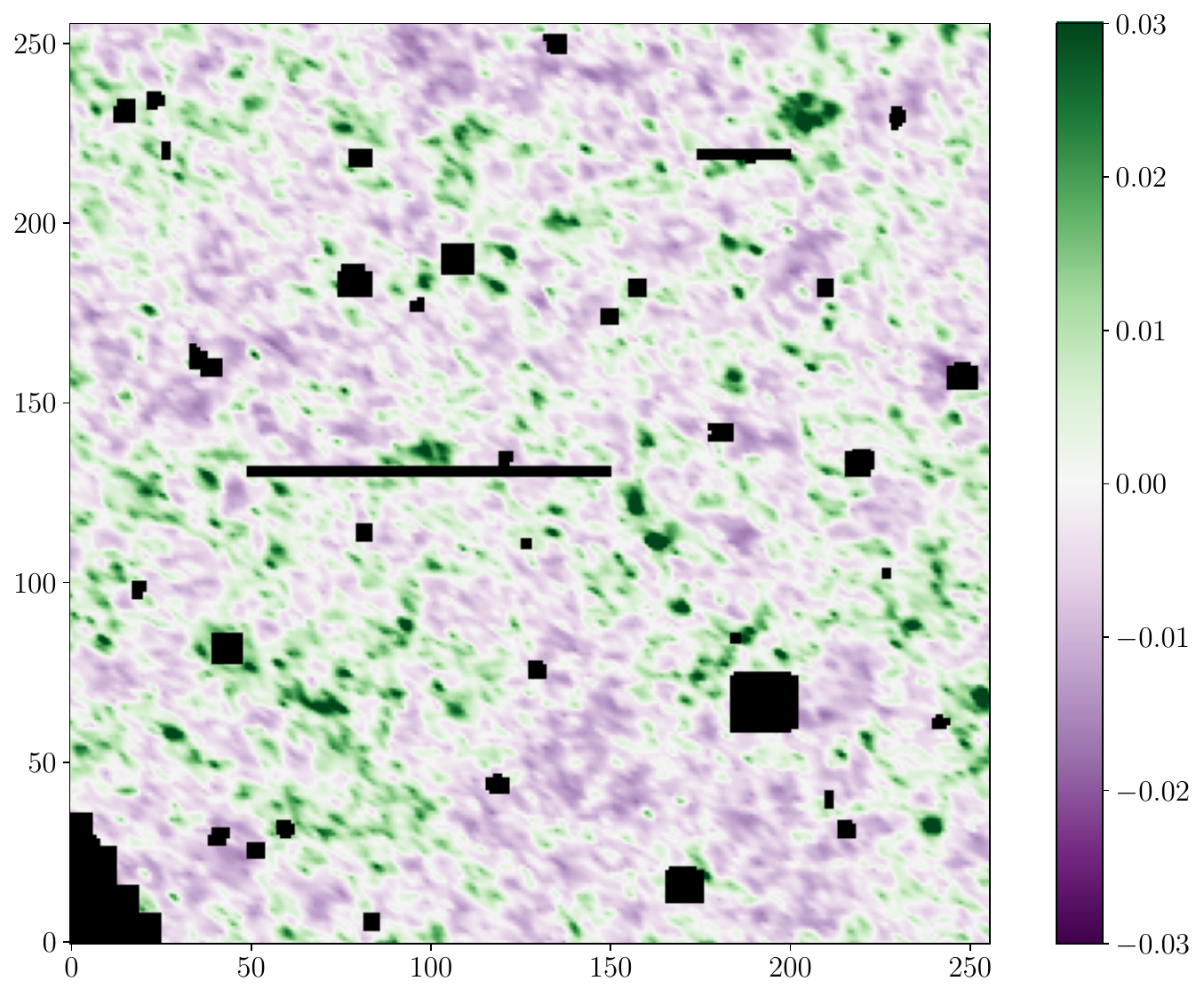}
    \caption{Mask applied to a 15 $\times$ 15 degree flat sky convergence map of redshift bin 0 at fiducial cosmology. The colourbar shows convergence of each pixel. We note that in practice the map is applied to shear map, only visualised here on convergence data.}
    \label{fig:mask}
\end{figure}

\hfill 

We apply a mask to our maps by hand to approximate the structure of real masks, covering approximately 2.8\% of the map, as seen in Figure \ref{fig:mask}. The regions in the mask do not correspond to specific simulations of real physical processes or artefacts, but rather we have manually created a mock mask matching the general shape of typical DES masks \citep{des_cat}.

After applying an inverse Kaiser-Squires transform to the C\textsc{osmo}G\textsc{rid}V1 convergence maps, we add the noise and mask to the resulting shear maps. A Gaussian kernel $G$ with a standard deviation of one pixel has to be applied to the masked shear map before reconstruction. To omit the masked pixels, we compute
\begin{equation}
    \frac{G * M}{G * B},
\end{equation}
where $M$ is the input map with the masked pixels set to zero and $B$ is a binary mask containing ones for unmasked pixels. This smoothing fills in gaps and decreases the resolution of the maps. This is a crucial step for Kaiser Squires, which only works for continuous fields.

\subsubsection{Applications of Reconstruction Methods}
For the Wiener filter reconstruction calculation of the signal matrix $\mathbf{S}$ in Eq.~\eqref{eq:wf}, we compute the 2D flat-sky power spectrum $P(k)$ with \texttt{CCL} \citep{ccl} using the cosmology of the fiducial convergence map 
and incorporating baryonic effects using \texttt{HMCode} \citep{bary_ccl,bary_param_vals}.

Splitting $k$ into its components $k_1$ and $k_2$ in Eq.~\eqref{eq:convergence}, $P(k)$ can be found in 2D space.  We relate the flat-sky 2D wavenumber $k$ to the angular multipole $\ell$ as
\begin{equation}
    \ell = \frac{2\pi}{bs}k,
\end{equation}
where $b$ is the box width in pixels and $s$ is the pixel size in radians. Using CCL, we then compute the angular power spectrum $C_\ell$ at the corresponding cosmology and perform a scalar transformation using
\begin{equation}
    P(k) = \frac{b}{s} C_\ell.
\end{equation}

The noise covariance matrix $\mathbf{N}$ in Eq.~\eqref{eq:wf} is a diagonal matrix with value $\sigma^2$ from Eq.~\eqref{eq:sigma}. It describes the statistics of the noise in the convergence map. 

We train and test the \texttt{DeepMass} model on realisations from a single (fiducial) cosmology and apply it to the nearby non-fiducial cosmologies. To test potential systematic uncertainties from this approach, we will also compare it to models trained directly on the non-fiducial data. 

\texttt{DeepMass} does not include a padding functionality in the model, so the output map does not have the same dimensions as the input map. As such, we pad input data to reduce edge effects on the signal. To account for any correlation in signal between different redshifts, we have extended \texttt{DeepMass} to a 3D model across all four redshift bins. We note sporadic issues where \texttt{DeepMass} yields NaN values. We then rerun the model at the same training cosmology.

\subsection{Observables}

The functional integrals from Eq.~\eqref{eq:v0},~\eqref{eq:v1}, and~\eqref{eq:v2} are converted to sums over the number of pixels $N$, so we can calculate MFs from the 2D reconstructed C\textsc{osmo}G\textsc{rid}V1 convergence fields:

\begin{equation}
    V_0(t_j) = \frac{1}{N} \sum_{i} \Theta(\kappa(\textbf{x}_i)-t),
    \label{v0_sum}
\end{equation}

\begin{equation}
    V_1(t_j) = \frac{1}{4N} \sum_{i} \Delta(\kappa(\textbf{x}_i)-t) \sqrt{\kappa_x^2 + \kappa_y^2},
    \label{v1_sum}
\end{equation}

\begin{multline}
    V_2(t_j) = \frac{1}{2\pi N} \sum_{i} \Delta(\kappa(\textbf{x}_i)-t) \\
    \left( \frac{2\kappa_x \kappa_y \kappa_{xy}-\kappa_x^2 \kappa_{yy}-\kappa_y^2 \kappa_{xx}}{\kappa_x^2 + \kappa_y^2} \right),
\label{v2_sum}
\end{multline}
where $\Delta$ is 1 when $\kappa(\textbf{x}_i)$ is between the thresholds $t_j$ and $t_{j+1}$ and 0 outside the range. Following \citet{grewal}, we evenly space the thresholds from $\mu - 3\sigma$ to $\mu + 3\sigma$, where $\mu$ is 0 and $\sigma$ is the field value standard deviation for each redshift in each cosmology. We differ slightly here, as in that work we chose thresholds for each individual map realisation. We note the thresholds are dynamic across reconstruction methods; we expect this to slightly improve our posterior contours compared to other approaches.

\subsection{Evaluating the Posterior}

We measure MFs from the pixels in the reconstructed posterior distributions of each method, then build a Fisher matrix $\mathrm{F}$ to evaluate the constraining power on cosmological parameters:

\begin{equation}
    \mathrm{F}_{ij} = \sum_{mn} \frac{\partial X_m}{\partial \theta_i} \mathrm{C}^{-1}_{mn} \frac{\partial X_n}{\partial \theta_j},
    \label{eq:fisher}
\end{equation}
where $X_m$ is the measured MF, $\theta_i$ is the cosmological parameter, and the covariance matrix $C_{mn}$ is $\langle (X_m - \overline{X}_m)(X_n - \overline{X}_n) \rangle$. The covariance is obtained from simulations at the fiducial cosmology. The Fisher derivatives are taken from the slope of the line of best fit through the three cosmologies in Table \ref{tab:delta_params}: minus parameter step size, fiducial, and plus parameter step size. The paired initial condition seeds decrease the noise in this calculation. To match DES Y3 constraining power, we divide the covariance matrix by the square root of the ratio of DES sky area (5000 sq deg) to our map size (15$\times$15 sq deg); this scale factor is 4.714 \citep{gatti_moments_2021}.

\begin{figure}[h]
    \centering
    \includegraphics[width=\columnwidth]{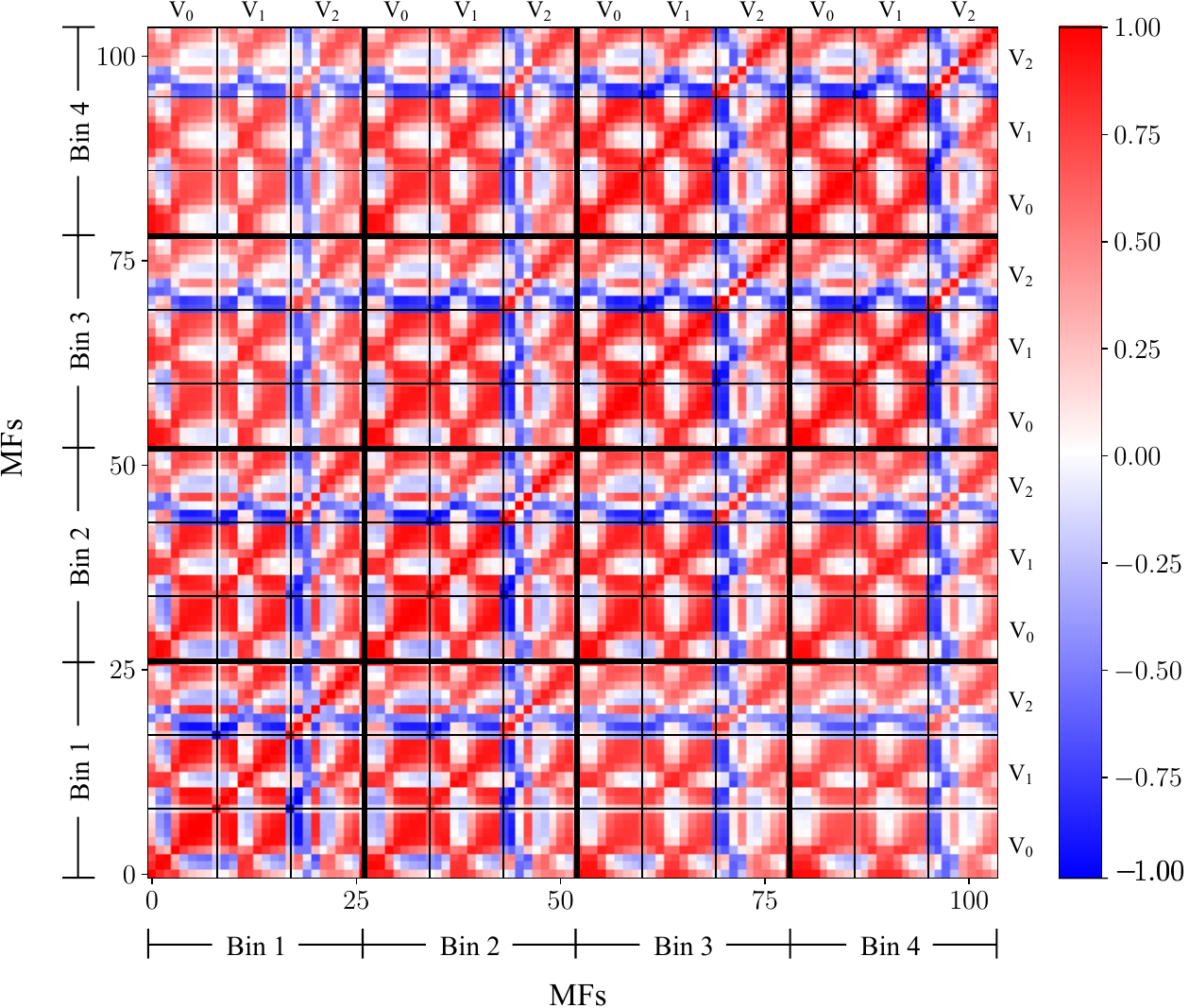}
    \caption{Correlation coefficient matrix of all MFs, which have been concatenated from all four redshift bins. There is strong correlation between the MFs across redshift. The boxes represent one functional, where there are three per bin (denoted by a thicker black line). The MFs have been measured from a true convergence map at fiducial cosmology.} 
    \label{fig:corr}
\end{figure}

The correlation matrix derived from the covariance matrix is shown in Figure \ref{fig:corr}.  While $V_0$ has some correlation with itself, it is anti-correlated with $V_1$ and $V_2$, which are highly correlated with each other. There is also strong correlation across redshift bins. We note the difference between this matrix and our previous MF correlation matrix in \citet{grewal} can be attributed to the thresholding methodology. 

With the center set at the fiducial cosmology, we follow the process in \citet{fisher} to generate the ellipse parameters and build the contours in Section \ref{section:Results}. We marginalise over the other parameters by inverting the Fisher matrix \textit{before} removing them.

\subsection{Figures of Merit}

We use two figures of merit (FOM) to measure the constraining power of different reconstruction methods: the lensing amplitude parameter $S_8$ and the inverse of the area of the $\Omega_\mathrm{m}$ - $\sigma_8$ contour ellipse. $S_8\equiv\sigma_8 \left(\frac{\Omega_\mathrm{m}}{0.3}\right)^{\alpha}$ (where $\alpha$=0.5) is perpendicular to the lensing degeneracy direction for $\Omega_\mathrm{m}$ and $\sigma_8$ for typical survey redshifts. It is thus the minor axis of the $\Omega_\mathrm{m}$ - $\sigma_8$ contours, and is usually the best constrained parameter combination.
 
Following \citet{euclid_metric}, we can transform the Fisher matrix in $\Omega_\mathrm{m}$ and $\sigma_8$ into the matrix for $\Omega_\mathrm{m}$ - $S_8$ with

\begin{equation}
    \mathrm{F} (\Omega_\mathrm{m},\mathrm{S}_8) = \mathrm{M^T} \mathrm{F}(\Omega_\mathrm{m},\sigma_8)\mathrm{M},
\end{equation}
where 

\begin{equation}
    \mathrm{M} = \begin{bmatrix}
    \dfrac{\partial\Omega_\mathrm{m}}{\partial\Omega_\mathrm{m}} & \dfrac{\partial\Omega_\mathrm{m}}{\partial \mathrm{S}_8} \\[10pt]
    \dfrac{\partial\sigma_8}{\partial\Omega_\mathrm{m}} & \dfrac{\partial\sigma_8}{\partial \mathrm{S}_8}
    \end{bmatrix} = \begin{bmatrix}
    1 & 0 \\
    -0.5\dfrac{\sigma_8}{\Omega_\mathrm{m}} & \left( \dfrac{0.3}{\Omega_\mathrm{m}}\right)^{0.5}
    \end{bmatrix}.
\end{equation}
Our code used to transform a Fisher matrix into a contour for a given set of cosmological parameters (including $\Omega_\mathrm{m}$ and $S_8$) is available here.\footnote{\url{https://github.com/nishagrewal/fisher_to_contour}}

In this work, we also aim to study the uncertainty in the \texttt{DeepMass} reconstruction. We quantify it with $\chi^2$ per pixel:

\begin{equation}
\chi^2 = \frac{1}{N_\mathrm{pixels}}\sum \left(\frac{X_\mathrm{recon} - X_\mathrm{true}}{\sigma}\right)^2,
\label{eq:chi}
\end{equation}
using the $\sigma$ defined in Eq.~\eqref{eq:sigma}. $X_\mathrm{true}$ is the true MF observable and $X_\mathrm{recon}$ is the reconstructed MF observable, fiducial or non-fiducial.

\section{Results}
\label{section:Results}

\begin{figure*}[t]
    \centering
    \includegraphics[width=0.6\textwidth]{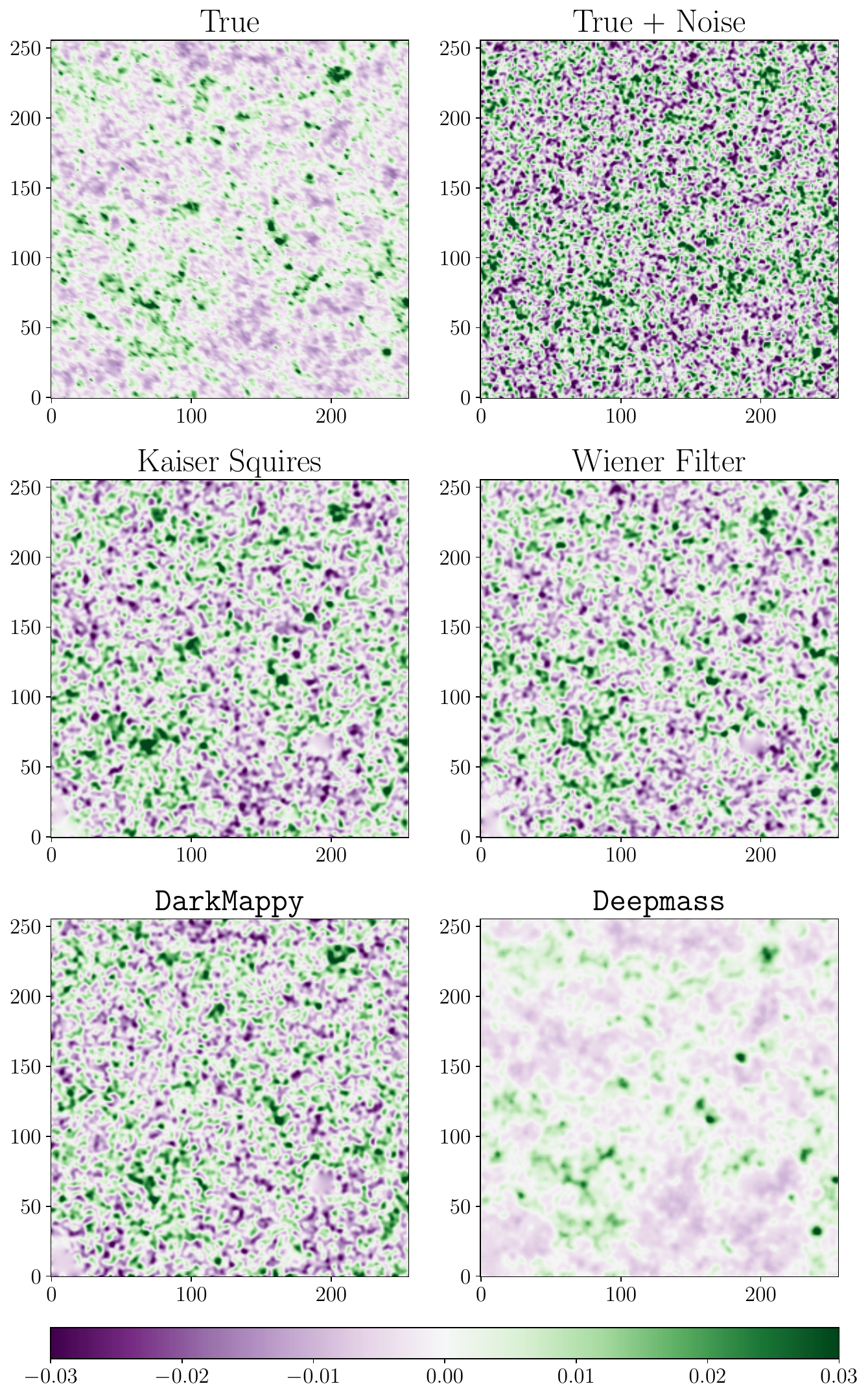}
    \caption{Visualising the output for true maps, true-plus-noise maps, and the four different reconstruction methods. 15 $\times$ 15 degree flat sky convergence maps of redshift bin 3 at fiducial cosmology. The shared colourbar shows convergence of each pixel. A Gaussian kernel with a standard deviation of one arcmin has been applied to the true map to achieve the smoothed map.}
    \label{fig:recon_compare}
\end{figure*}

Figure \ref{fig:recon_compare} shows the true, true smoothed, and reconstructed convergence maps for the four methods: Kaiser Squires, Wiener filter, \texttt{DarkMappy}, and \texttt{DeepMass}. As expected, peaks in the KS and WF maps have been smoothed out because small scales are noise-dominated; however, \texttt{DarkMappy} looks similarly noise-dominated despite it being a non-linear method. \texttt{DeepMass} appears most like the truth convergence map, which is evidence that it has denoised the convergence signal; this is consistent with predictions. While \texttt{DeepMass} takes more time to train the model than the other methods do to run, once the model is generated, \texttt{DeepMass} reconstructs convergence maps very quickly.

\begin{figure*}[h]
    \centering
    \includegraphics[width=\textwidth]{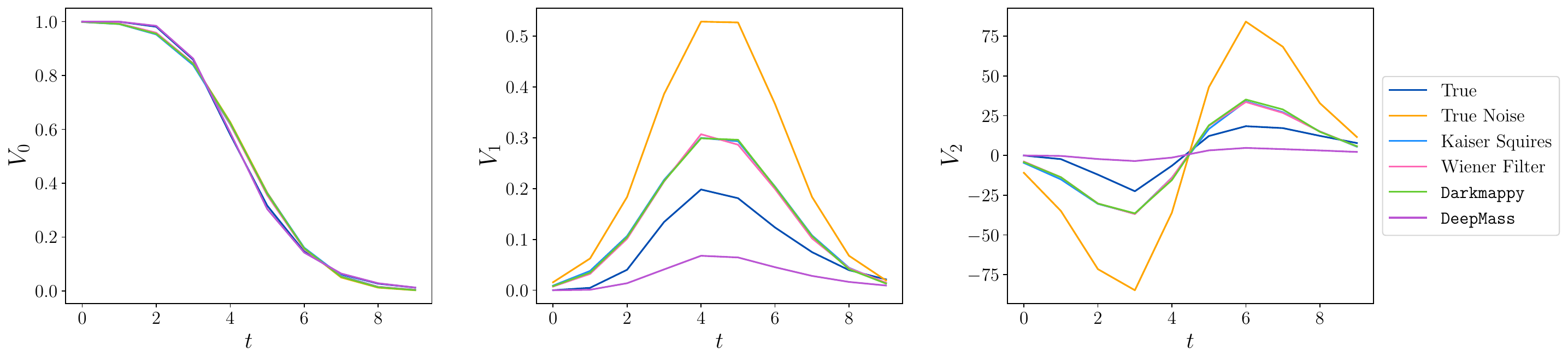}
    \caption{The first three MFs, $V_0$, $V_1$, and $V_2$, calculated from fiducial convergence maps. The curves shown are for true signal and true-plus-noise, along with the four reconstructed signals visualised in Figure \ref{fig:recon_compare}: Kaiser Squires, Wiener filter, \texttt{DarkMappy}, and \texttt{DeepMass}. The $t$ values on the curve correspond to an excursion set, and the three MFs on the $y$-axis describe the perimeter, area, and curvature of the excursion set of the convergence.}
    \label{fig:mf_nm}
\end{figure*}

Since MFs are not linear, their total value does not come from summing the MFs from noise and signal. Here we have a separate analysis with signal and noise to study the effects of the latter. The Minkowski functional outputs for true, true-plus-noise, KS, WF, \texttt{DarkMappy}, and \texttt{DeepMass} are compared in Figure \ref{fig:mf_nm}. The MF curves have the same shape, which is the typical shape for a convergence field. The distributions are not expected to be exactly the same, and here we see differing amplitudes for different methods corresponding to the noise level in the field. We do not expect KS to reduce the noise in any way, but some of the noise in KS has been smoothed out. As such, KS is not as noisy as true-plus-noise. Figure \ref{fig:mf_nm} shows KS, WF, and \texttt{DarkMappy} are comparable, while \texttt{DeepMass} has the lowest variation in amplitude due to noise suppression after smoothing.

\begin{figure*}[h]
\centering
    \begin{tabular}{cc}
        \includegraphics[width=\columnwidth]{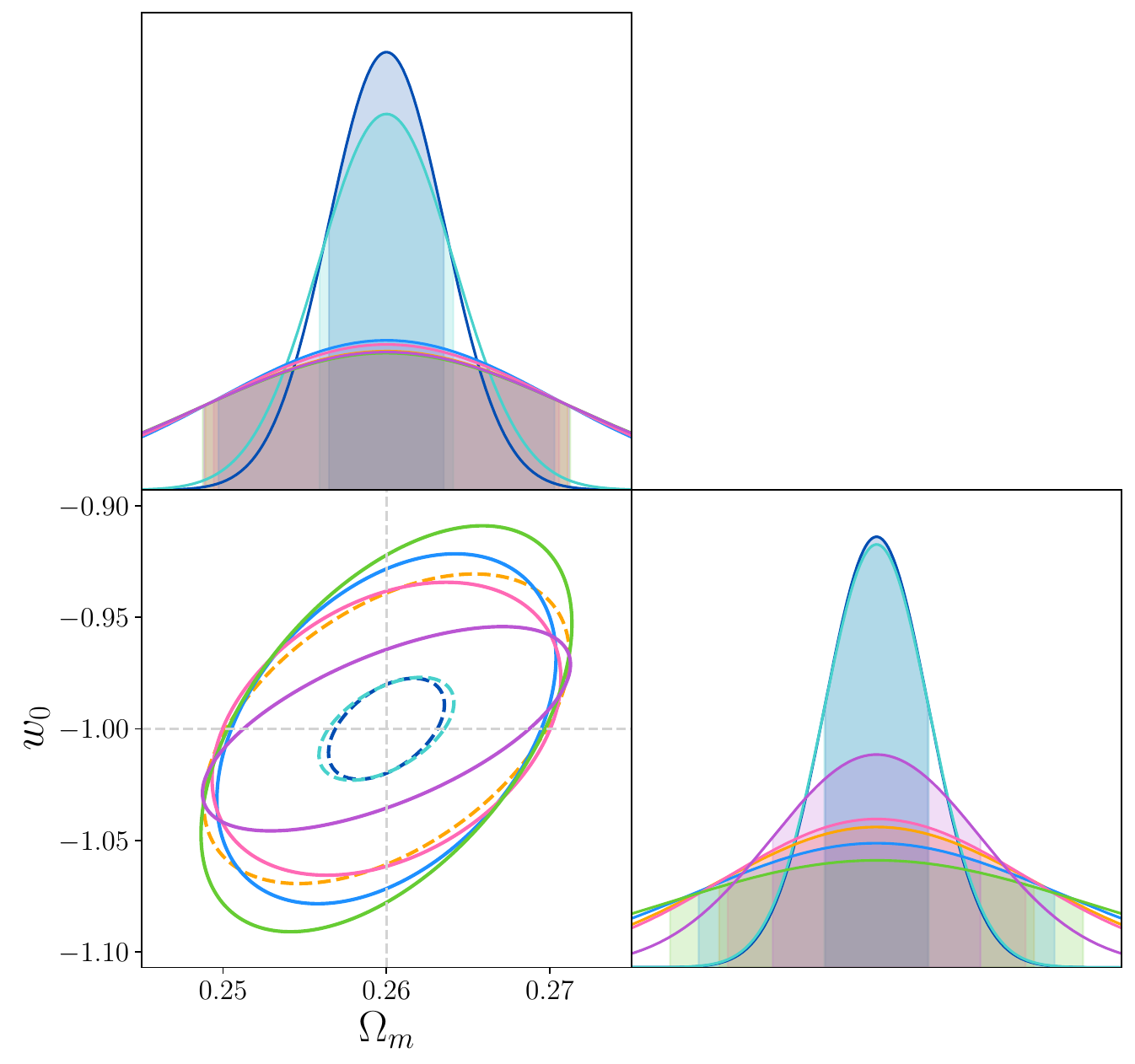} & \\
        \includegraphics[width=\columnwidth]{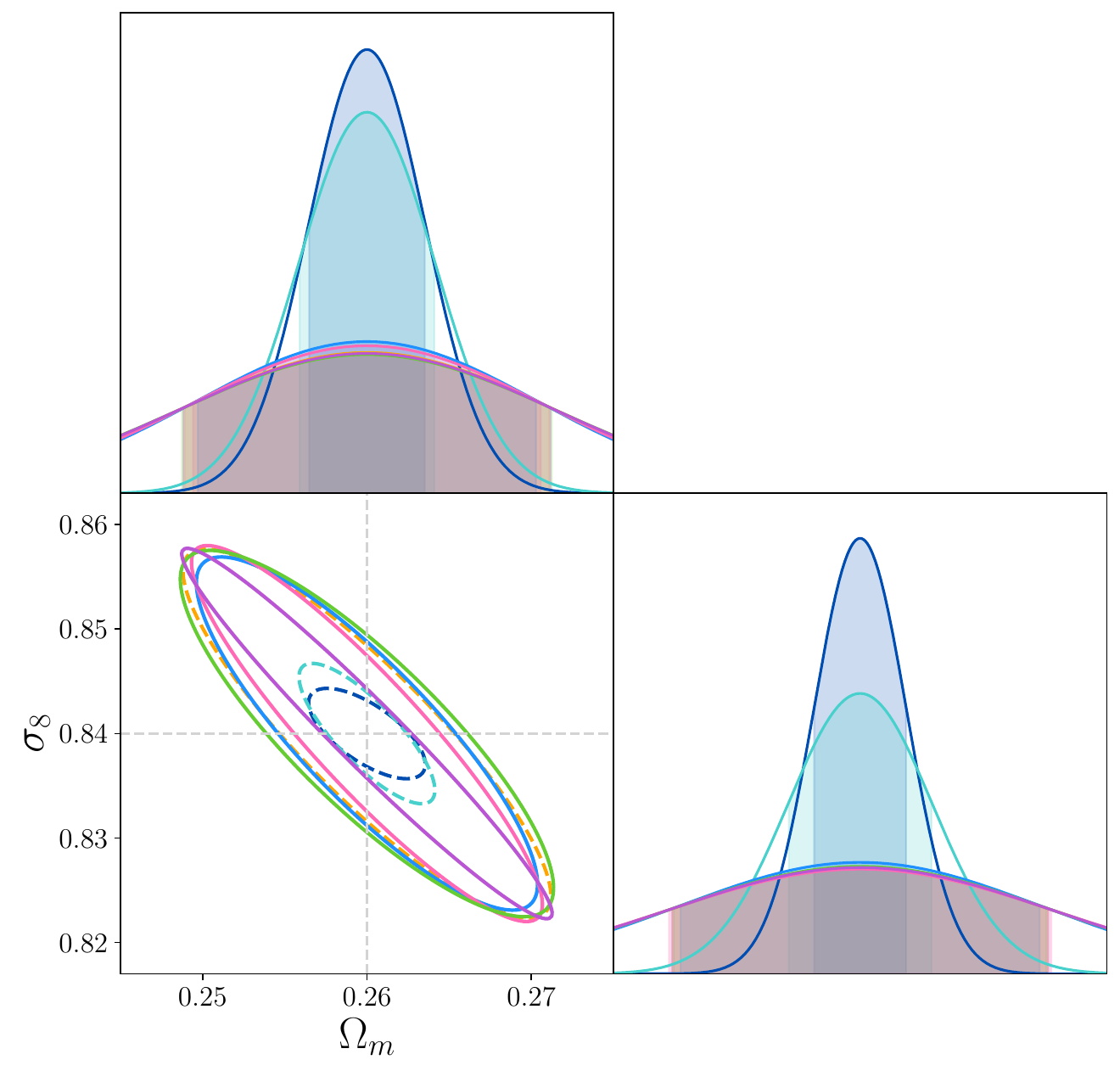} & \includegraphics[width=\columnwidth]{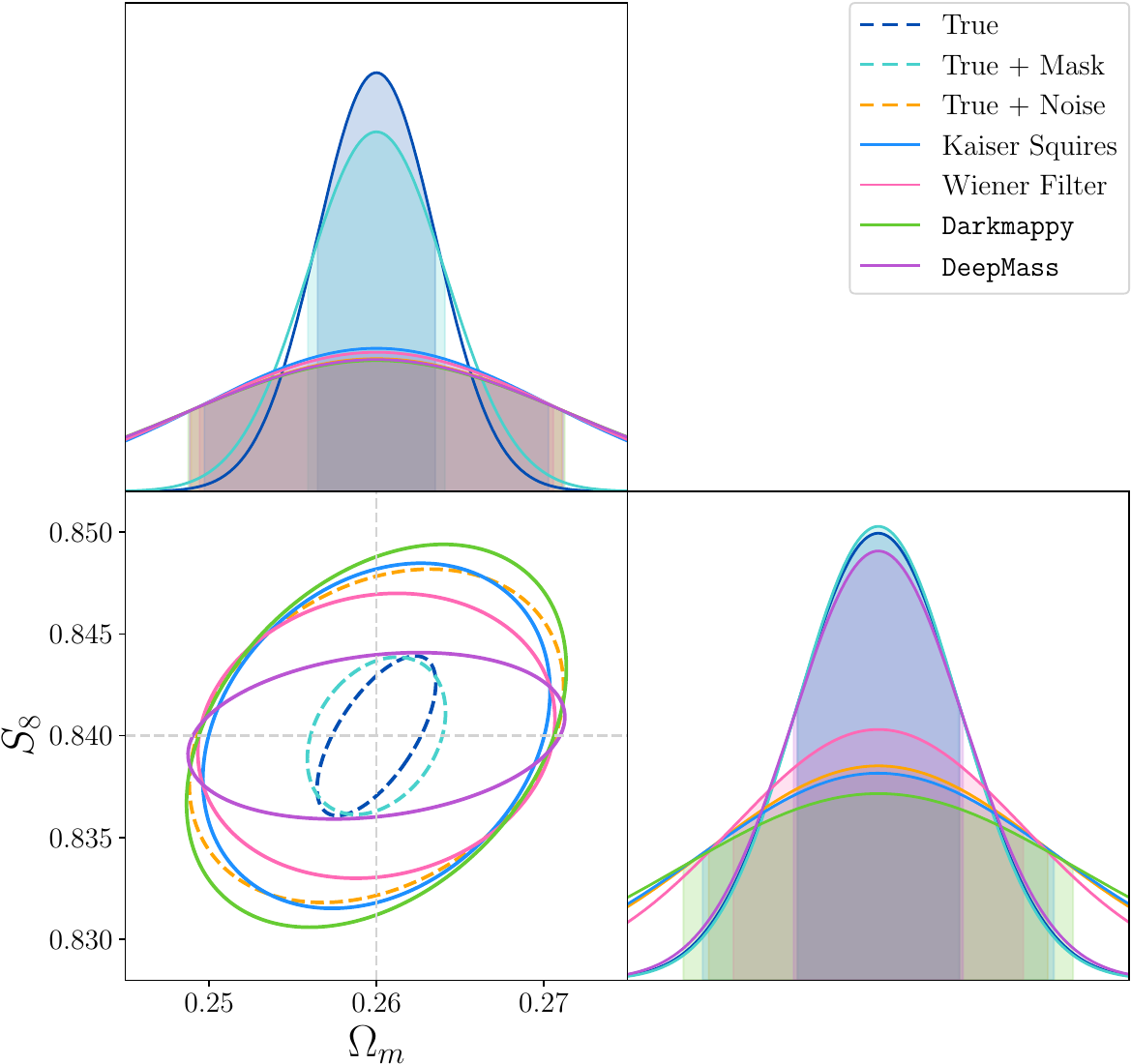} \\
    \end{tabular}
    \caption{$w$CDM Fisher constraints for true signal, true-plus-mask, true-plus-noise, KS reconstruction, WF reconstructions, \texttt{DarkMappy} reconstruction, and \texttt{DeepMass} reconstruction. The top left plot shows constraints on $\Omega_\mathrm{m}$ vs $w_0$, the bottom left plot shows constraints on $\Omega_\mathrm{m}$ vs $\sigma_8$, and the bottom right plot shows constraints on $\Omega_\mathrm{m}$ vs $S_8$. KS, WF, and \texttt{DarkMappy} are comparable at these scales, while \texttt{DeepMass} shows significantly tighter constraints, implying the model has effectively denoised the signal.}
\label{fig:contours}
\end{figure*}

We marginalise over all cosmological and baryonic parameters in Table \ref{tab:delta_params} in our analysis except for the two used in our Fisher matrix. Figure \ref{fig:contours} shows the contour comparison in the reconstruction methods for $\Omega_\mathrm{m}$ - $\sigma_8$, $\Omega_\mathrm{m}$ - $S_8$, and $\Omega_\mathrm{m}$ - $w_0$. The contour directions are typical for convergence constraints. Figures of merit are shown in Table \ref{tab:fom} with the standard deviation of $S_8$ in the first column and the inverse of the $\Omega_\mathrm{m}$ - $\sigma_8$ contour area in the second column.

\begin{table}[h]
  \centering
  \begin{tabular}{lll}
    Method & $\sigma_{S_8}$ & 1/$E_A$ \\
    \midrule
    True & 0.00392 & 28328.11 \\ [2pt]
    True + Mask & 0.00387 & 19445.44 \\ [2pt]
    True + Noise & 0.00819 & 3378.91 \\ [2pt]
    Kaiser Squires & 0.00848 & 3484.66 \\ [2pt]
    Wiener Filter & 0.00700 & 3994.15 \\ [2pt]
    \texttt{DarkMappy} & 0.00940 & 2967.18 \\ [2pt]
    \texttt{Deepmass} & 0.00409 & 6606.95 \\ [2pt]
    \bottomrule
  \end{tabular}
  \caption{\textup{Figures of merit for different reconstruction methods. $E_A$ is the ellipse area of the $\Omega_\mathrm{m}$ - $\sigma_8$ contour.}}
  \label{tab:fom}
\end{table}

\texttt{DeepMass} outperforms the other reconstruction methods by roughly 50$\%$ and has constraints closest to the true contour. KS, WF, and \texttt{DarkMappy} have similar constraints within a 30\% interval of each other, and are comparable to true-plus-noise, which shows adding noise doubles the size of the contour. We note the choice of threshold range can affect the size of the contours. As mentioned previously, KS is not expected to perform as well as true-plus-noise, but some of the noise has been smoothed out. We can see in Figure \ref{fig:contours} that the step of masking the data has a much smaller impact than the step of adding noise to the fields, the latter doubling the size of the parameter constraints. We use standard value 0.5 for $\alpha$ in the $\sigma_{S_8}$ calculation, which corresponds to the width perpendicular to the direction of the true-plus-mask contour. We also include the second column for inverse ellipse area to more completely quantify constraining power of cosmological parameters, and we see that the true map has the largest value, followed by true-plus-mask. On the whole, $\Omega_\mathrm{m}$ is better constrained than other cosmological parameters. We can see the same trends in the $\Omega_\mathrm{m}$ - $S_8$ and $\Omega_\mathrm{m}$ - $w_0$ planes. We investigate baryonic parameters in Appendix \ref{section:Appendix}. 

\begin{figure}[h]
    \centering
    \includegraphics[width=\columnwidth]{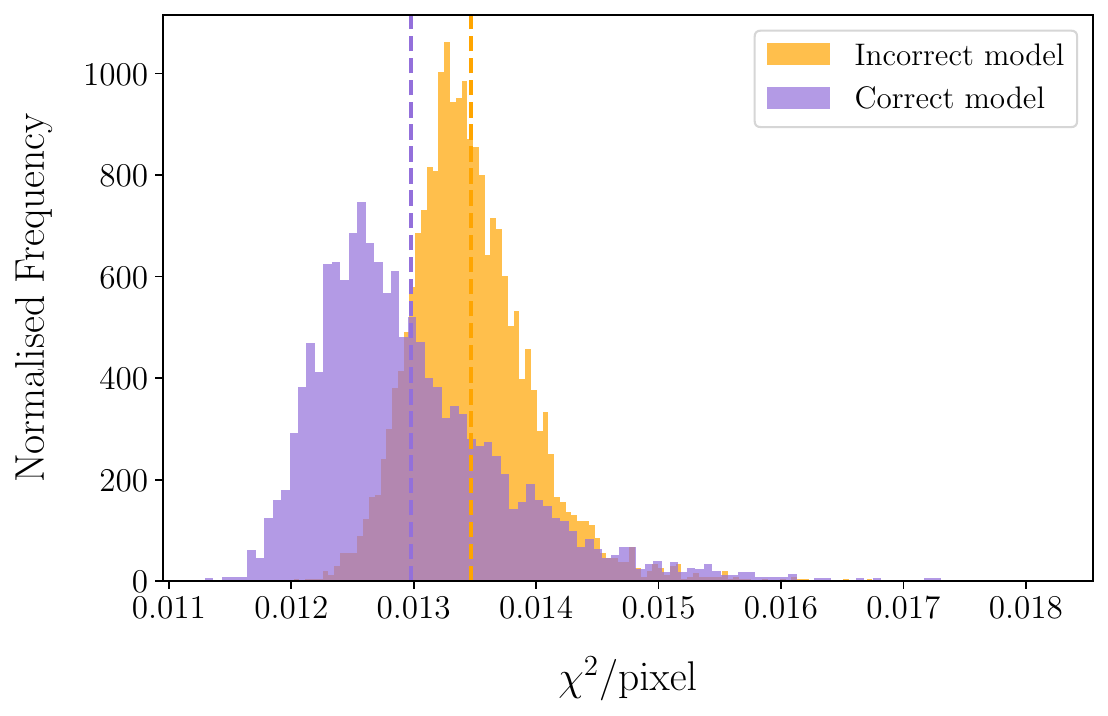}
    \caption{Normalised $\chi^2$/pixel comparison of \texttt{DeepMass} reconstruction using the correct model versus the incorrect model. The correct model (purple) gives a lower $\chi^2$ value than the incorrect model (orange), indicating that choice of model is important. We measure a 3.8\% uncertainty in the cosmology choice in \texttt{DeepMass} training, which has been calculated as the percent change in the difference in the means of the two $\chi^2$ histograms.}
    \label{fig:chi}
\end{figure}

\begin{figure*}[h]
    \centering
    \includegraphics[width=\textwidth]{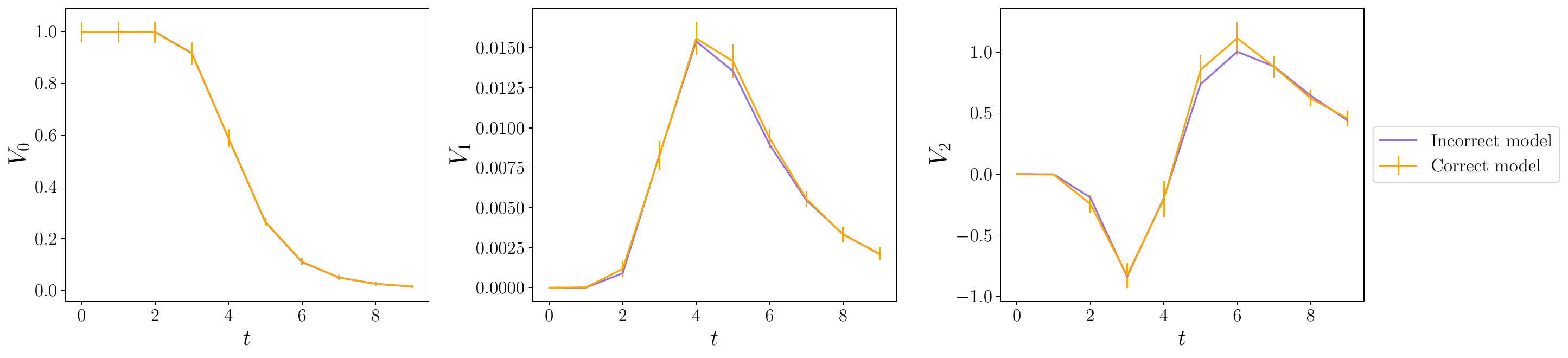}
    \caption{MFs for the two \texttt{DeepMass} outputs in Figure \ref{fig:chi}. Error bars are shown on the output from the correct model.}
    \label{fig:chi_mf}
\end{figure*}

Using Eq.~\eqref{eq:chi}, we compare $\chi^2$ values per pixel to measure the effect of training \texttt{DeepMass} at a (slightly) wrong cosmology. Figure \ref{fig:chi} shows a comparison of \texttt{DeepMass} outputs for the fiducial and non-fiducial models (-$\Delta\Omega_\mathrm{m}$) applied to non-fiducial data (-$\Delta\Omega_\mathrm{m}$). Each histogram displays the $\chi^2$ values across 5000 output maps. The incorrect model histogram illustrates the distinction between the fiducial model and true data with -$\Delta\Omega_\mathrm{m}$, while the correct model histogram depicts differences for the non-fiducial model on the same input data. As expected, the histogram for the correct model has a lower average $\chi^2$ value. The quantification of uncertainty from the reconstruction is determined as the difference between the means of the two histograms: 3.8\%. While this seems relatively small, it leads to changes in the MFs comparable to cosmologically-induced changes, as seen in Figure \ref{fig:chi_mf}. As this can only worsen the constraining power of \texttt{DeepMass}, which is already close to the ideal constraints from truth data, it cannot have had a significant effect on our results here, as we focus on the reconstruction precision (not accuracy). However, this demonstrates that the selection of training model cosmologies is critical for future analysis accuracy. Tests in \citet{deepmass} used the truth cosmologies for optimal training, but analyses of real data will need thoughtful and perhaps iteratively-chosen training samples and careful emulators to interpolate between them. 

\section{Conclusion}
\label{section:Conclusion}

We want to use the most precise reconstructed convergence maps in order to best capture the original signal. In this way, measuring statistics from such detailed reconstructed maps enables improved measurement of cosmological parameters. In this paper we investigate four non-parametric mass mapping methods developed to reconstruct convergence from shear: Kaiser Squires, Wiener filter, \texttt{DarkMappy}, and \texttt{DeepMass}. We smooth masked and noisy shear maps and reconstruct convergence maps using the different methods. KS requires smoothing because it can only take a continuous field as input. We then measure Minkowski functionals from the reconstructed convergence maps and perform a Fisher analysis to make contours. We compare constraints on $\Omega_\mathrm{m}$, $\sigma_8$, $S_8$, and $w_0$ (see Appendix \ref{section:Appendix} for baryonic parameter analysis).

The contour plots for different combinations of cosmological parameters all show the same results for different reconstruction methods. \texttt{DeepMass} outperforms the rest of the methods by a significant degree, demonstrating the model is a successful denoiser and an effective tool for reconstructing convergence from shear. Beyond a 50$\%$ improvement in constraining power, the model is also computationally efficient. 

As \texttt{DeepMass} denoises the Wiener filter (WF) output convergence, the smoothing in WF implies a certain scale where the signal dominates for \texttt{DeepMass}. \texttt{DeepMass} has the capability to go beyond this with either higher resolution maps, lower noise, or a different initialisation that does not rely on WF.

We also measure the uncertainty from using different cosmologies in the \texttt{DeepMass} model training. By comparing $\chi^2$ values of true data MFs and MFs from data trained on the correct model or the same data trained on an incorrect model, we can quantify the impact of cosmological choices on model performance with the difference in average $\chi^2$ values. This is about 3.8\%, indicating the choice of training model cosmology has an impact on reconstruction precision. 

This difference aligns closely with observed variations in average MF \texttt{DeepMass} outputs from data trained on the correct or incorrect model. Both effects are comparable with the change in cosmology for $\Omega_m$. These findings demonstrate a strong dependence of \texttt{DeepMass} on the cosmology of the training dataset, emphasising the need for the consideration of input cosmology in model training to ensure accurate and reliable results. While a larger training set could improve the signal reconstruction, a simulation-based inference framework is perhaps a safer solution since it can more explicitly incorporate these uncertainties.

The remaining methods, KS, WF, and \texttt{DarkMappy} are comparable at the scales we use here. \texttt{DarkMappy} was expected to outperform KS and WF. One reason it did not could be that it was designed for cluster analysis and may be tuned for such scenarios rather than our wide-field tests; alternative choices of its parameters and wavelet dictionary could improve its performance. Another could be the resolution of the field. In this paper, we have used C\textsc{osmo}G\textsc{rid}V1 simulations with DES Y3-like redshift bins and noise, which have led to a more limited resolution. We expect an improvement in resolution with less noisy survey conditions like for LSST or in future LSST-like forecasts, as well as with newer simulations.

Additionally, the choice of threshold levels can influence contour sizes. Here we use dynamic thresholds, where the range is set by the mean and standard deviation of the field for each cosmology, redshift, and reconstruction method. Using fixed thresholds more carefully tuned to each case has less variability. This can lead to smaller contours, but the thresholds are less consistently useful for different inputs like redshift bins, reconstruction methods, and simulation conditions like noise level.

We use Minkowski functionals to evaluate the different reconstruction methods, as they are quick to measure, unbiased, and typically resilient to systematic uncertainties. However, there are many other (combinations of) statistics that could be used to probe the underlying dark matter distribution. The two-point correlation function measures two-point information, and other higher order statistics may yield different results and show more distinction between the contours. Likewise, we have chosen four reconstruction methods here, representative of the different approaches to mass mapping. Other methods may be as or more precise.

In future analysis, using different (or multiple statistics) could lead to better information about how the different methods perform. Additionally, using lower noise conditions like LSST and other reconstruction methods could lead to stronger differentiation among methods. 

We finally note that mass mapping for higher-order statistics is an ideal problem for a community challenge like the ones presented in \citet{tomo_challenge} and \citet{n5k}. This would both focus authors of reconstruction methods on useful metrics for upcoming surveys and incentivise them to optimise parameters and choices for a specific scenario.

\section{Acknowledgements}
We thank Tomasz Kacprzak for help with the C\textsc{osmo}G\textsc{rid}V1 simulations. NG thanks Harry Rendell-Bhatti for the useful discussions. TT acknowledges funding from the Swiss National Science Foundation under the Ambizione project PZ00P2\textunderscore193352. Results in this paper made use of many software packages, including \texttt{Numpy}, \texttt{Scipy}, and \texttt{CCL} \citep{numpy,scipy,ccl}.

\bibliographystyle{aasjournal}
\bibliography{main}

\vspace*{500px}
\appendix

\section{Effects of Baryons}
\label{section:Appendix}

\subsection{Comparing Fisher calculations for Baryonic Parameter Constraints}

\begin{figure}[ht]
        \includegraphics[width=0.5\linewidth]{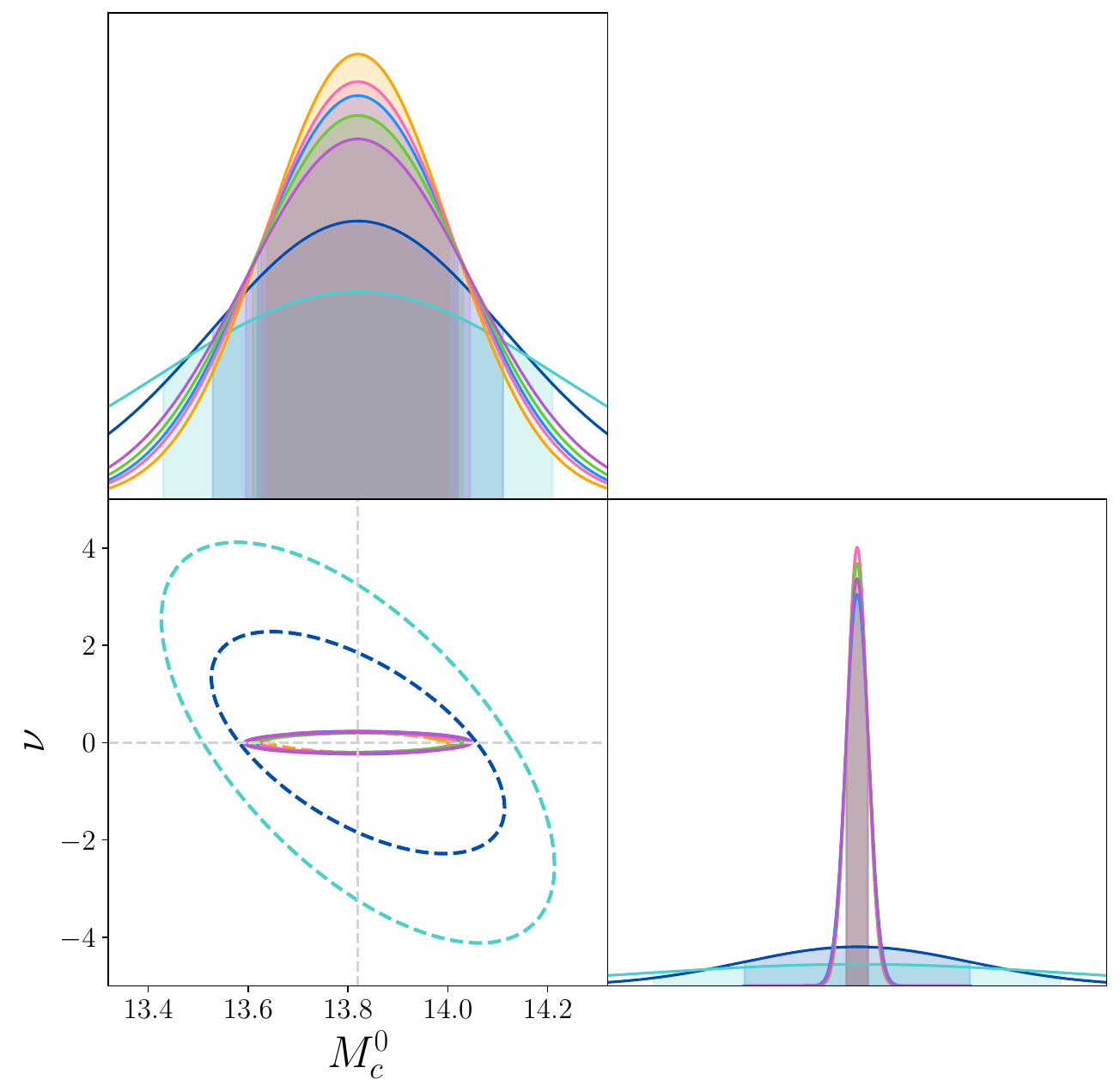}
        \includegraphics[width=0.5\linewidth]{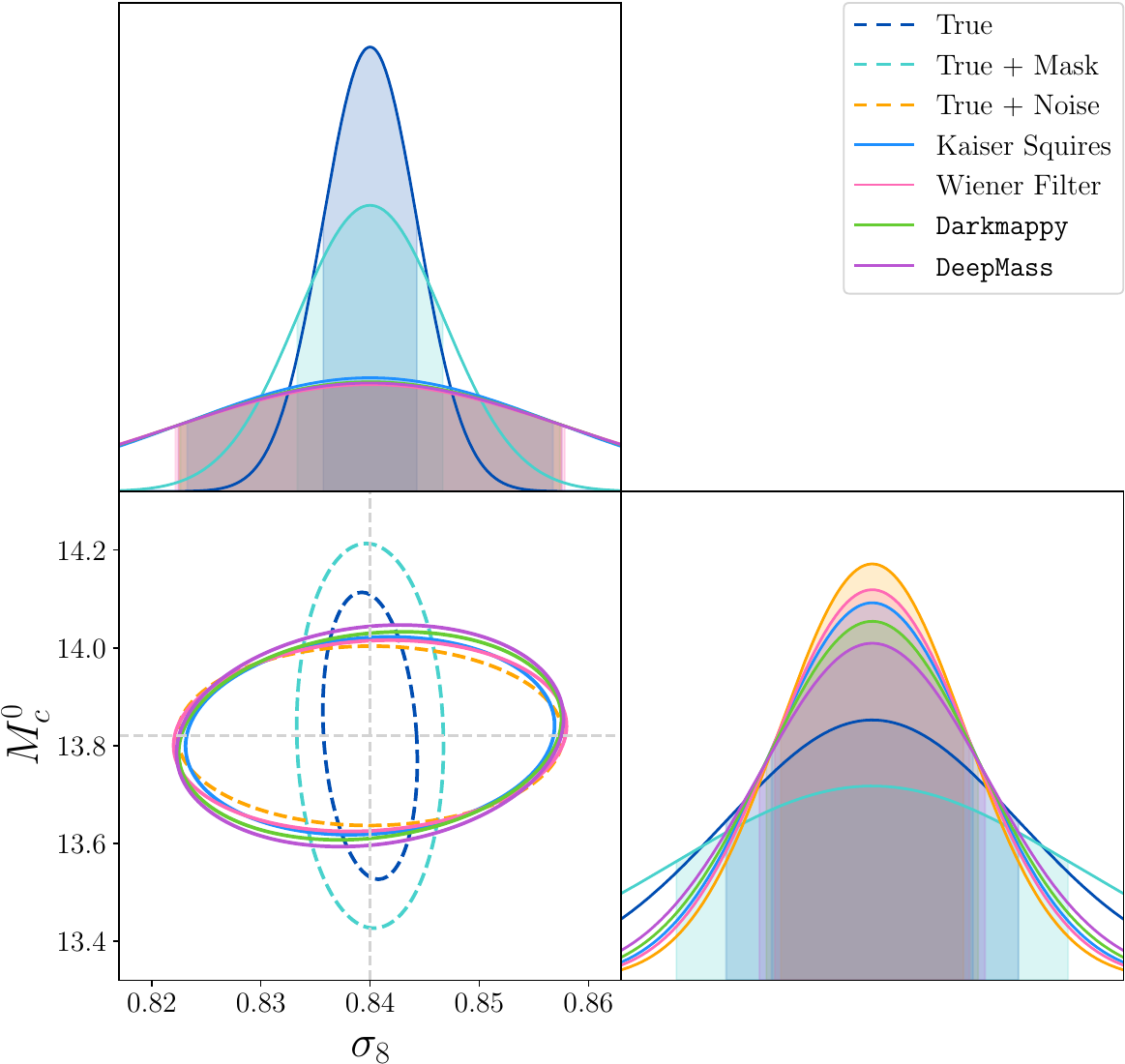}
\caption{Contours for different baryonic parameter combinations. The same four reconstruction methods are compared, which produce similar constraints, with \texttt{DeepMass} performing the best.}
\label{fig:ks_fisher_and_bary_contour}
\end{figure}

Here we explore the effects of baryons in our simulations and analysis. Figure \ref{fig:ks_fisher_and_bary_contour} shows contour plots for baryonic parameters. The left plot shows the $M^0_\mathrm{c}-\nu$ plane, and the right plot shows the $\sigma_8-M^0_\mathrm{c}$ plane. The results are unexpected, as the true maps should have the tightest constraints. We would expect smoothing and noise in the reconstruction process to lead to a loss of information, especially for small-scales where we expect to see baryonic effects. These contours are not reliable. To investigate this discrepancy, we study the baryonic effects at each step of the analysis.

\subsection{Baryonic non-fiducial Convergence Maps}

\begin{figure}[h]
    \centering
    \includegraphics[width=\textwidth]{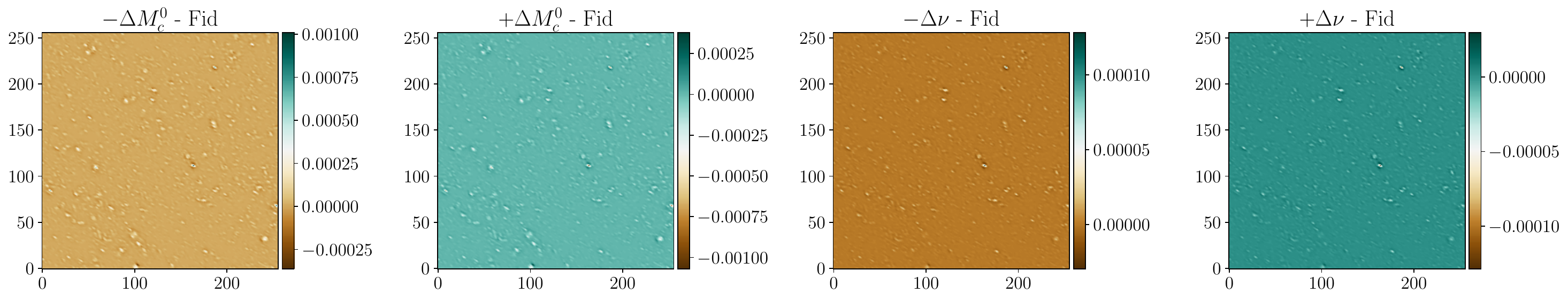}
    \caption{The difference in the non-fiducial and fiducial convergence maps for baryon parameters $M_\mathrm{c}^0$ and $\nu$.}
    \label{fig:bary_map_diff}
\end{figure}

We first compare the true baryonic and fiducial convergence maps to visualise the effect of a change in the baryonic parameters $M_\mathrm{c}^0$ and $\nu$. Figure \ref{fig:bary_map_diff} shows the difference between the non-fiducial $\pm\Delta M_\mathrm{c}^0$ and $\pm\Delta \nu$ baryonic convergence maps and the fiducial convergence maps. We can see that the change in signal is very small compared to the original convergence value. This is roughly two orders of magnitude smaller than the change for nonbaryonic parameters.

\subsection{Minkowski functionals for Baryonic Parameters}
\begin{figure}[ht]
    \centering
    \begin{minipage}{\textwidth}
        \centering
        \includegraphics[width=\textwidth]{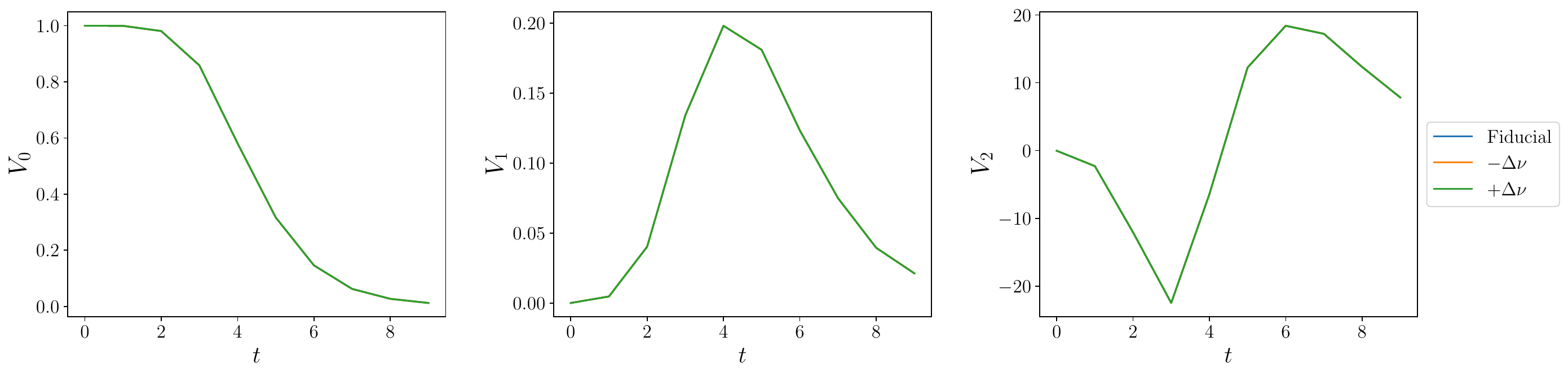}
    \end{minipage}
    \hfill
    \begin{minipage}{\textwidth}
        \centering
        \includegraphics[width=\textwidth]{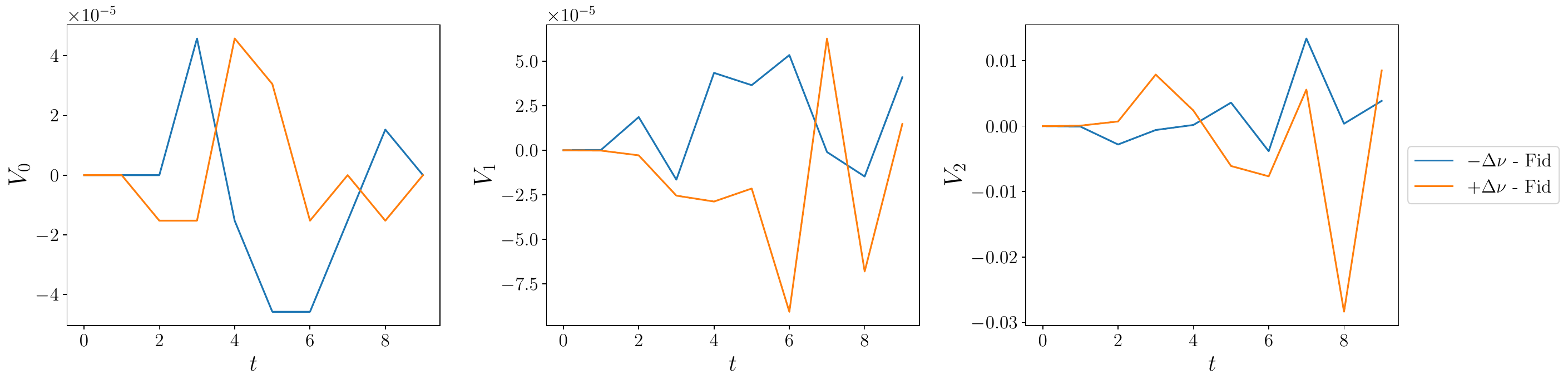}
    \end{minipage}
    \caption{MFs at different values of $\nu$ (top row). Difference in MFs of $\Delta\nu$ and fiducial maps (bottom row).}
    \label{fig:bary_nu_mf}
\end{figure}

\begin{figure}[ht]
    \centering
    \begin{minipage}{\textwidth}
        \centering
        \includegraphics[width=\textwidth]{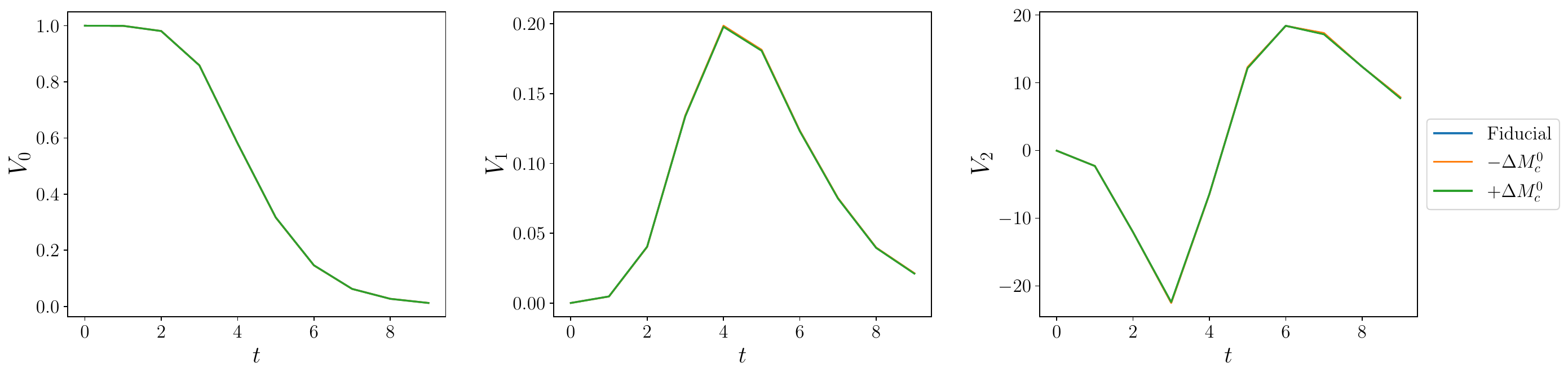}
    \end{minipage}
    \hfill
    \begin{minipage}{\textwidth}
        \centering
        \includegraphics[width=\textwidth]{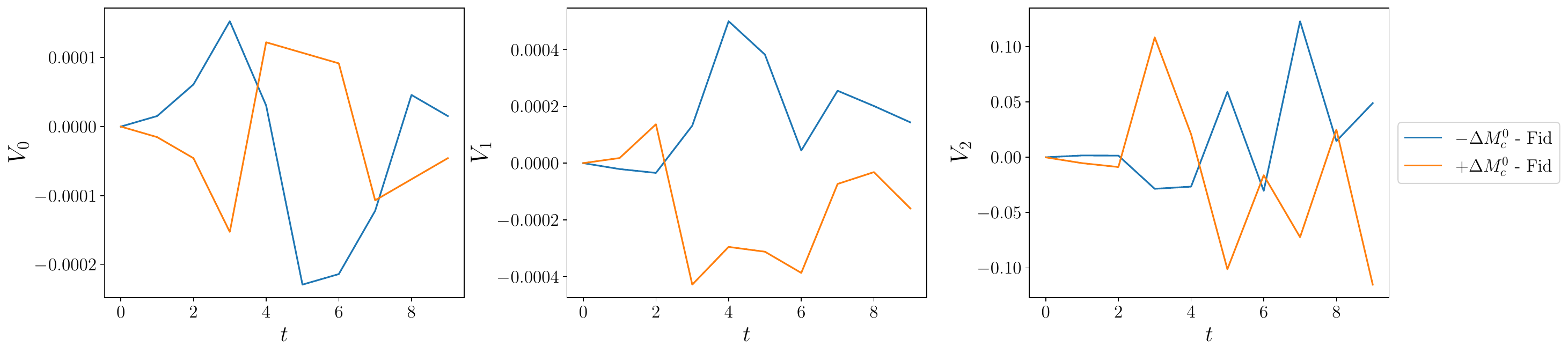}
    \end{minipage}
    \caption{MFs at different values of $M_\mathrm{c}^0$ (top row). Difference in MFs of $\Delta M_\mathrm{c}^0$ and fiducial maps (bottom row).}
    \label{fig:bary_Mc_mf}
\end{figure}

We next compare the Minkowski functional observables measured from the baryonic and fiducial convergence maps. In Figures \ref{fig:bary_nu_mf} and \ref{fig:bary_Mc_mf}, the functionals at different baryon cosmologies are shown to be nearly identical. This is not the case for the cosmological parameters we constrain in this paper; an example of variation in MF observables for $\Omega_\mathrm{m}$ is shown in Figure \ref{fig:mf_Om}. Here, any difference in the MFs falls within the noise regime. This may be because the $\Delta$ parameter step size is too small or the pixel size is too large to visualise or measure baryonic effects. These changes in step size are small enough that numerical noise dominates true Fisher matrix contours.

\section{Minkowski functionals for Non-Baryonic Parameters}

\begin{figure}[h]
    \centering
    \includegraphics[width=\textwidth]{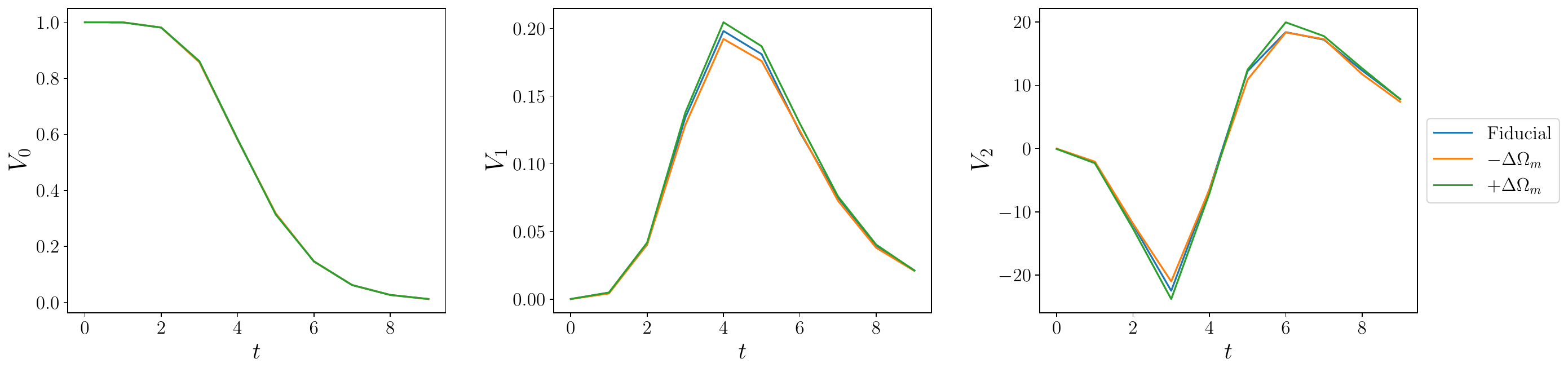}
    \caption{MFs at different values of $\Omega_\mathrm{m}$; the change is comparable to the variation from training cosmology in Figure \ref{fig:chi_mf}.}
    \label{fig:mf_Om}
\end{figure}

For non-baryonic parameters, we see the expected change in the Minkowski functionals that corresponds to a change in cosmology. Figure \ref{fig:mf_Om} shows MFs for $\pm\Delta\Omega_\mathrm{m}$ as well as the fiducial value for $\Omega_\mathrm{m}$.

\end{document}